\documentclass[12pt]{article}

\usepackage{epsfig}

\newcommand{\sect}[1]{\setcounter{equation}{0}\section{#1}}
\renewcommand{\theequation}{\arabic{section}.\arabic{equation}}

\def\be{\begin{equation}}
\def\ee{\end{equation}}
\def\ba{\begin{eqnarray}}
\def\ea{\end{eqnarray}}

\title{{\bf Brane New World}}
\author{
S.W. Hawking\thanks{email: S.W.Hawking@damtp.cam.ac.uk},
T. Hertog\thanks{Aspirant FWO-Vlaanderen; email: T.Hertog@damtp.cam.ac.uk} and
H.S. Reall\thanks{email: H.S.Reall@damtp.cam.ac.uk}
\\ \\ DAMTP \\ Centre for Mathematical Sciences \\ University of
Cambridge \\ Wilberforce Road, Cambridge CB3 0WA, UK. 
\\ \\ Preprint DAMTP-2000-25}

\date{March 7, 2000}

\begin{document}

\maketitle

\begin{abstract}

We study a Randall-Sundrum cosmological scenario consisting
of a domain wall in anti-de Sitter space with a strongly coupled large
$N$ conformal field theory living on the wall. The AdS/CFT
correspondence allows a fully quantum mechanical treatment of this
CFT, in contrast with the usual treatment of matter fields in inflationary
cosmology. The conformal anomaly of the CFT provides an effective 
tension which leads to a de Sitter geometry for the domain wall. This
is the analogue of Starobinsky's four dimensional model of anomaly
driven inflation. Studying this model in a Euclidean
setting gives a natural choice of boundary conditions at the
horizon. We calculate the graviton correlator using the Hartle-Hawking
``No Boundary'' proposal and analytically continue to Lorentzian
signature. We find that the CFT strongly suppresses metric perturbations on
all but the largest angular scales. This is true independently of how
the de Sitter geometry arises, i.e., it is also true for four
dimensional Einstein gravity. Since generic matter would be
expected to behave like a CFT on small scales, our results suggest
that tensor perturbations on small scales are far smaller than
predicted by all previous calculations, which have neglected 
the effects of matter on tensor perturbations. 

\end{abstract}

\sect{Introduction}

Randall and Sundrum (RS) have suggested \cite{rs2} that four 
dimensional  gravity may be recovered in the presence of an infinite
fifth dimension provided that we live on a domain wall embedded in
anti-de Sitter space (AdS). Their linearized analysis showed that
there is a massless bound state of the graviton associated with 
such a wall as well as a continuum of massive Kaluza-Klein modes.  
More recently, linearized analyses have examined the spacetime
produced by matter on the domain wall and concluded that it is in
close agreement with four dimensional Einstein gravity \cite{gt,gkr}.

\begin{figure}
\centerline{\psfig{file=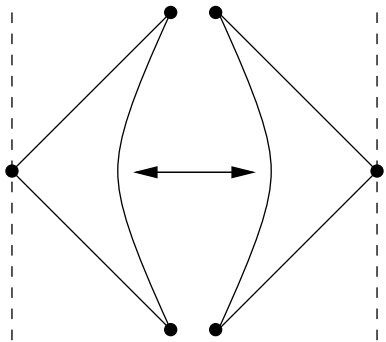,width=2.in}}
\caption{Carter-Penrose diagram of anti-de Sitter space with a flat domain
wall. The dotted line denotes timelike infinity and the arrows denote 
identifications. The heavy dots denote points at infinity. Note that
the Cauchy horizons intersect at infinity.}
\label{fig:flat}
\end{figure}

RS used horospherical coordinates based on slicing AdS into flat
hypersurfaces. These horospherical coordinates break down at the
horizons shown in figure \ref{fig:flat}.
An issue that has not received much attention so far is the role of
boundary conditions at these Cauchy horizons in AdS.
With stationary perturbations, one can impose the boundary conditions
that the horizons remain regular. Indeed, without this boundary
condition the solution for stationary perturbations is not well
defined. Even for non-perturbative  departures from the RS solution,
like black holes, one can  impose the boundary condition that the
AdS horizons remain regular \cite{wallbh,ehm1,gt,ehm2,6dbh}.  
Non-stationary perturbations on the domain wall,
however, will give rise to  gravitational waves that cross the
horizons. This will tend to focus the  null geodesic generators of the
horizon, which will mean that they  will intersect each other on some
caustic. Beyond the caustic, the null geodesics will not lie in the
horizon. However, null geodesic generators of the future event horizon
cannot have a future endpoint \cite{he} and so the endpoint must lie
to the past. We conclude that if the past and future horizons remain
non-singular when perturbed\footnote{
It has been shown that the KK modes of RS give rise to singular
horizons \cite{ppwave}.} 
(as required for a well-defined boundary condition) 
then they must intersect at a finite distance from the wall.
By contrast, the past and future horizons don't intersect in the
RS ground state but go off to infinity in AdS. 

The RS horizons are like the horizons of extreme black
holes. When considering perturbations of black holes, one normally
assumes that radiation can flow across the future horizon but that
nothing comes out of the past horizon. This is because the past
horizon isn't really there, and should be replaced by the collapse
that formed the black hole.
To justify a similar boundary condition on the Randall-Sundrum past
horizon, one needs to consider the initial conditions of the
universe. 

\begin{figure}
\centerline{\psfig{file=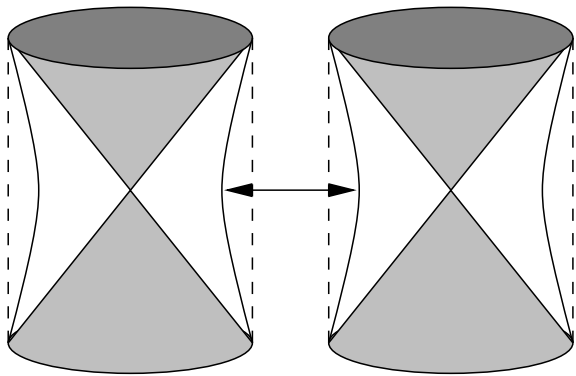,width=3.in}}
\caption{Anti-de Sitter space with a de Sitter domain wall. AdS is
drawn as a solid cylinder, with the boundary of the cylinder (dashed line)
representing timelike infinity. The light cone shown is the
horizon. The arrows denote identifications.}
\label{fig:deS}
\end{figure}

The main contender for a theory of initial conditions is the
``no boundary'' proposal\footnote{
Other approaches to quantum cosmology in the RS model have been 
discussed in \cite{gs,qcos}. Boundary conditions motivated by a
Euclidean approach were also used in \cite{gkr} for a flat domain
wall.} \cite{nbp} that the quantum state of the universe is given by a
Euclidean path integral over compact metrics.
The simplest way to implement this proposal
for the Randall Sundrum idea is to take the Euclidean version of the
wall to be a four sphere at which two balls of $AdS_5$ are joined
together.
In other words, take two balls in $AdS_5$, and glue them
together along their four sphere boundaries. The result is
topologically a five sphere, with a delta function of curvature on a
four dimensional domain wall separating the two hemispheres. If one
analytically continues to Lorentzian signature, one obtains a four
dimensional de Sitter hyperboloid, embedded in Lorentzian anti de Sitter 
space, as shown in figure \ref{fig:deS}. 
The past and future RS horizons, are replaced by the past
and future light cones of the points at the centres of the two
balls. Note that the past and future horizons now intersect each
other and are non extreme, which means they are stable to
small perturbations. A perfectly spherical Euclidean domain wall will give
rise to a four dimensional Lorentzian universe that expands
forever in an inflationary manner\footnote{ 
Such inflationary brane-world solutions have been
studied in \cite{dddw,kaloper,nihei,kim2,gs}. For a discussion 
of other cosmological aspects of the RS model, see \cite{cgrt} and 
references therein.}.

In order for a spherical domain wall solution to exist, the tension of
the wall must be larger than the value assumed by RS, who had a flat
domain wall. We shall assume that matter on the wall increases its
effective tension, permitting a spherical solution. 
In section \ref{sec:CFT}, we consider a strongly coupled
large $N$ CFT on the domain
wall. On a spherical domain wall, the conformal anomaly of the CFT
increases the effective tension of the domain
wall, making the spherical solution possible. The Lorentzian geometry
is a de Sitter universe with the conformal anomaly driving
inflation\footnote{
A similar idea was recently discussed within the context of
renormalization group flow in the AdS/CFT correspondence
\cite{noz}. However, in the case the CFT was the CFT dual to the bulk AdS
geometry, not a new CFT living on the domain wall.}
, an idea introduced long ago by Starobinsky \cite{star1}.

The no boundary proposal allows one to calculate
unambiguously the graviton correlator on the domain wall. 
In particular, the Euclidean path integral itself uniquely
specifies the allowed fluctuation modes, 
because perturbations that have
infinite Euclidean action are suppressed in the path integral.
Therefore, in this framework, there is no need to impose by hand 
an additional, external 
prescription for the vacuum state for each perturbation
mode. In addition, 
the AdS/CFT correspondence 
allows a fully quantum mechanical treatment of the CFT, in
contrast with the usual classical treatment of matter fields in inflationary
cosmology.

Finally, we analytically continue the Euclidean correlator into the
Lorentzian region, where it describes the spectrum of 
quantum mechanical vacuum 
fluctuations of the graviton field on an inflating domain
wall with conformally invariant matter living on it. 
We find that the quantum loops of the large $N$
CFT give spacetime a rigidity that strongly suppresses metric
fluctuations on small scales. Since any matter would be expected to
behave like a CFT at small scales, this result probably extends to any
inflationary model with sufficiently many matter fields. It has long
been known that matter loops lead to short distance modifications of
gravity. Our work shows that these modifications can lead to observable
consequences in an inflationary scenario.

Although we have carried out our calculations for the RS model, we
shall show that results for four dimensional Einstein gravity
coupled to the CFT can be recovered by taking the domain wall to be 
large compared with the AdS scale. Thus our conclusion that metric
fluctuations are suppressed holds independently of the RS scenario. 

The spherical domain wall considered in this paper analytically
continues to a Lorentzian de Sitter universe that inflates
forever. However, Starobinsky \cite{star1} showed that the conformal
anomaly driven de Sitter phase is unstable to evolution into a matter
dominated universe. If such a solution could be obtained from a
Euclidean instanton then it would have an $O(4)$ symmetry group,
rather than the $O(5)$ symmetry of a spherical instanton. We shall
study such models for both the RS model and four
dimensional Einstein gravity in a separate paper.

The AdS/CFT correspondence \cite{malda,gkp,witten} provides an
explanation of the RS behaviour\footnote{
This was first pointed out in unpublished remarks of Maldacena and Witten.}
 \cite{gubser}.  It relates the
RS model to an equivalent four dimensional theory
consisting of general relativity coupled to a strongly interacting
conformal field theory and a logarithmic correction.
Under certain circumstances, the effects of
the CFT and logarithmic term are negligible and pure gravity is 
recovered. We review this correspondence in section \ref{sec:RSCFT}.

In section \ref{sec:CFT} we present our calculation of the graviton
correlator on the instanton and demonstrate how the result is
continued to Lorentzian signature.
Section \ref{sec:conclude} contains our conclusions and some
speculations. This paper also includes two appendices which
contain technical details that we have omitted from the text.

\sect{Randall-Sundrum from AdS/CFT}

\label{sec:RSCFT}

The AdS/CFT correspondence \cite{malda,gkp,witten} relates
IIB supergravity theory in $AdS_5\times S^5$ to a ${\cal N} = 4$ 
$U(N)$ superconformal
field theory. If $g_{YM}$ is the coupling constant of this theory then
the 't Hooft parameter is defined to be $\lambda = g_{YM}^2 N$. The
CFT parameters are related to the supergravity parameters by \cite{malda}
\be
 l = \lambda^{1/4} l_s,
\ee
\be
\label{eqn:Ndef}
 \frac{l^3}{G} = \frac{2N^2}{\pi},
\ee
where $l_s$ is the string length, $l$ the AdS radius and $G$ the {\it
five} dimensional Newton constant. Note that $\lambda$ and $N$ must be
large in order for stringy effects to be small. 
The CFT lives on the conformal boundary of
$AdS_5$. The correspondence takes the following form:
\be
\label{eqn:corres}
 Z[{\bf h}] \equiv \int d[{\bf g}] \exp (-S_{grav}[{\bf g}])
 = \int d[\phi] \exp (-S_{CFT}[\phi;{\bf h}]) \equiv \exp(-W_{CFT}[{\bf
 h}]),
\ee
here $Z[{\bf h}]$ denotes the supergravity partition function in
$AdS_5$. This is given by a path integral over all metrics in
$AdS_5$ which induce a given conformal equivalence class of metrics
${\bf h}$ on the conformal boundary of $AdS_5$. The correspondence
relates this to the generating functional $W_{CFT}$ of connected Green's
functions for the CFT on this boundary. This functional is given by a
path integral over the fields of the CFT, denoted schematically by
$\phi$. Other fields of the supergravity theory can be included on the
left hand side; these act as sources for operators of the CFT on the
right hand side. 

A problem with equation \ref{eqn:corres} as it stands is that the
usual gravitational action in AdS is divergent, rendering the path
integral ill-defined. A procedure for solving this problem was
developed in \cite{witten,tseytlin,hensken1,hensken2,bal,ejm,kraus}. 
First one brings the boundary
into a finite radius. Next one adds a finite number of counterterms 
to the action in order to render it finite as the boundary is moved
back off to
infinity. These counterterms can be expressed solely in 
terms of the geometry of the boundary. The total gravitational action
for $AdS_{d+1}$ becomes
\be
 S_{grav} = S_{EH} + S_{GH} + S_1 + S_2 + \ldots.
\ee
The first term is the usual Einstein-Hilbert action\footnote
{We use a positive signature metric and a curvature convention for
which a sphere has positive Ricci scalar.} with a negative
cosmological constant:
\be
 S_{EH} = -\frac{1}{16 \pi G} \int d^{d+1} x \sqrt{g} \left(R +
\frac{d(d-1)}{l^2}\right)
\ee
the overall minus sign arises because we are considering a Euclidean
theory. The second term in the action is the Gibbons-Hawking boundary
term, which is necessary for a well-defined variational problem
\cite{gh}:
\be
 S_{GH} = -\frac{1}{8 \pi G} \int d^d x \sqrt{h} K,
\ee
where $K$ is the trace of the extrinsic curvature of the
boundary\footnote{Our convention is the following. 
Let $n$ denotes the outward unit normal to the
boundary. The extrinsic curvature is defined as $K_{\mu\nu} = h_{\mu}
^{\rho} h_{\nu}^{\sigma} \nabla_{\rho} n_{\sigma}$, where
$h_{\mu}^{\nu}=\delta_{\mu}^{\nu}-n_{\mu} n^{\nu}$ projects quantities
onto the boundary.} and $h$ the determinant of the induced metric. 
The first two counterterms are given by the following
\cite{hensken2,bal,ejm,kraus} (we use the results of \cite{kraus}
rotated to Euclidean signature) 
\be
 S_1 = \frac{d-1}{8\pi G l} \int d^d x \sqrt{h},
\ee
\be
 S_2 = \frac{l}{16\pi G (d-2)} \int d^d x \sqrt{h} R,
\ee
where $R$ now refers to the Ricci scalar of the boundary metric. The
third counterterm is 
\be
\label{eqn:CT3div}
 S_3 = \frac{l^3}{16\pi G (d-2)^2 (d-4)} \int d^d x \sqrt{h}
\left(R_{ij} R^{ij} - \frac{d}{4(d-1)} R^2 \right),
\ee
where $R_{ij}$ is the Ricci tensor of the boundary metric and boundary
indices $i,j$ are raised and lowered with the boundary metric
$h_{ij}$. This expression is ill-defined for $d=4$, which is the case
of most interest to us. With just the first two counterterms, the
gravitational action exhibits logarithmic divergences
\cite{tseytlin,hensken1,hensken2} so a third term is needed. This term cannot
be written solely in terms of a polynomial in scalar invariants of the
induced metric and curvature tensors;
it makes explicit reference to the cut-off (i.e. the finite radius to
which the boundary is brought before taking the limit in which it
tends to infinity). The form of this term is the same as
\ref{eqn:CT3div} with the divergent factor of $1/(d-4)$ replaced by
$\log (R/\rho)$, where $R$ measure the boundary radius and $\rho$ is
some finite renormalization length scale.

Following \cite{gubser}, we can now use the AdS/CFT correspondence to
explain the behaviour discovered by Randall and Sundrum. The
(Euclidean) RS model has the following action:
\be
 S_{RS}=S_{EH}+S_{GH}+2 S_1+S_m.
\ee
Here $2 S_1$ is the action of a domain wall with tension $(d-1)/(4\pi
G l)$. The final term is the action for any matter present on the domain
wall. The domain wall tension can cancel the effect of the bulk
cosmological constant to produce a flat domain wall. However, we are interested
in a spherical domain wall so we assume that the matter on the wall
gives an extra contribution to the effective tension. We shall discuss
a specific candidate for the matter on the wall later on. The wall
separates two balls $B_1$ and $B_2$ of $AdS$. 

We want to study quantum fluctuations of the metric on the domain
wall. Let ${\bf g}_0$ denote the five dimensional 
background metric we have just
described and ${\bf h}_0$ the metric it induces on the wall.
Let ${\bf h}$ denote a metric perturbation on the
wall. If we wish to calculate correlators of ${\bf h}$ on the domain
wall then we are interested in a path integral of the form\footnote{
In principle, we should worry about gauge fixing and ghost
contributions to the gravitational action. A convenient gauge to use
in the bulk is transverse traceless gauge. We shall only deal
with metric perturbations that also appear transverse and traceless on the
domain wall. The gauge fixing terms vanish for such perturbations and
the ghosts only couple to these perturbations at higher orders.}
\be
 \langle h_{ij}(x) h_{i'j'}(x') \rangle = \int d[{\bf h}] Z[{\bf h}]
 h_{ij}(x) h_{i'j'}(x'), 
\ee 
where
\ba
 Z[{\bf h}] &=& \int_{B_1 \cup B_2} d[\delta{\bf g}] d[\phi] \exp(-S_{RS}[{\bf
 g}_0 + \delta {\bf g}]) \nonumber \\
 &=& \exp(-2S_1[{\bf h}_0 + {\bf h}])  \\
 &{}& \times  \int_{B_1 \cup B_2} d[\delta{\bf
 g}] d[\phi] \exp(-S_{EH}[{\bf g}_0 + \delta {\bf g}] - S_{GH}[{\bf
 g}_0 + \delta {\bf g}] - S_m[\phi;{\bf h}_0 + {\bf h}]), \nonumber
\ea
$\delta {\bf g}$ denotes a metric perturbation in the bulk that
approaches ${\bf h}$ on the boundary and $\phi$ denotes the matter
fields on the domain wall. The integrals in the two balls
are independent so we can replace the path integral by
\ba
 Z[{\bf h}] &=& \exp(-2S_1[{\bf h}_0 + {\bf h}]) \left( \int_B d[\delta{\bf g}]
\exp(- S_{EH}[{\bf g}_0 + \delta {\bf g}] - S_{GH}[{\bf g}_0 + \delta
{\bf g}]) \right)^2 \nonumber \\
 &{}& \times \int d[\phi] \exp(- S_m[\phi;{\bf h}_0 + {\bf h}]),
\ea
where $B$ denotes either ball. We now take $d=4$ and use the AdS/CFT
correspondence \ref{eqn:corres} to replace the path integral over
$\delta {\bf g}$ by the generating functional for a conformal field
theory:
\ba
 \int_B && d[\delta {\bf g}] \exp(- S_{EH}[{\bf g}_0 + \delta {\bf g}] -
 S_{GH}[{\bf g}_0 + \delta {\bf g}]) = {} \nonumber \\
 &{}& \exp(-W_{RS}[{\bf h}_0 + {\bf
 h}] + S_1[{\bf h}_0 + {\bf h}] + S_2[{\bf h}_0 + {\bf h}] + S_3[{\bf
 h}_0 + {\bf h}]),
\ea
we shall refer to this CFT as the RS CFT since it arises as the dual
of the RS geometry. It has gauge group $U(N_{RS})$, where $N_{RS}$ is
given by equation \ref{eqn:Ndef}. Strictly speaking, we are using an
extended form of the AdS/CFT conjecture, which asserts that
supergravity theory in a finite region of AdS is dual to a CFT on the
boundary of that region with an ultraviolet cut-off related to the
radius of the boundary\footnote{Evidence in support of this extended
version of the duality was given in \cite{susswit}.}. 
The path integral for the metric perturbation becomes
\be
 Z[{\bf h}] = \exp(-2W_{RS}[{\bf h}_0 + {\bf h}] + 2S_2[{\bf h}_0 +
 {\bf h}] + 2S_3[{\bf h}_0 + {\bf h}]) \int d[\phi] \exp(-S_m[\phi;{\bf
 h}_0 + {\bf h}]).
\ee
The RS model has been replaced by a CFT and a coupling to matter
fields and the domain wall metric given by the action
\be
 -2S_2[{\bf h}_0 + {\bf h}] - 2S_3[{\bf h}_0 + {\bf h}] +
 S_m[\phi;{\bf h}_0 + {\bf h}].
\ee
The remarkable feature of this expression is that the term $-2S_2$ is
precisely the (Euclidean) Einstein-Hilbert action for {\it four
dimensional} gravity with a Newton constant given by the RS value
\be
 G_4 = G/l.
\ee
Therefore the RS model is equivalent to four dimensional gravity
coupled to a CFT with corrections to gravity coming from the third
counter term. This explains why gravity is trapped to the domain wall.

\medskip

At first sight this appears rather amazing. We started off with a
quite complicated five dimensional system and have argued that it is
dual to four dimensional Einstein gravity with some corrections and
matter fields. However in order to use this description, we have to
know how to calculate with the RS CFT. At present, the only way we
know of doing this is via AdS/CFT, i.e., going back to the five
dimensional description. The point of the AdS/CFT argument is to explain why
the RS ``alternative to compactification'' works and also to explain
the origin of the corrections to Einstein gravity in the RS
model. Note that if the matter on the domain wall dominates the RS CFT
and the third counterterm then these can be neglected and a purely
four dimensional description is adequate.

\sect{CFT on the Domain Wall}

\label{sec:CFT}

\subsection{Introduction}

Long ago, Starobinsky studied the cosmology of a universe containing
conformally coupled matter \cite{star1}. CFTs generally exhibit a conformal
anomaly when coupled to gravity (for a review, see
\cite{duff}). Starobinsky gave a de Sitter solution in which the
anomaly provides the cosmological constant. By analyzing
homogeneous perturbations
of this model, he showed that the de Sitter phase is unstable but
could be long lived, eventually decaying to a FRW cosmology. 

In this section we will consider the RS analogue of Starobinsky's model by
putting a CFT on the domain wall. On a spherical domain wall, the
conformal anomaly provides the extra tension required to satisfy the 
Israel equations. It is appealing to choose the new CFT to be a
${\cal N}=4$ superconformal field theory because then
the AdS/CFT correspondence makes calculations relatively
easy\footnote{
We emphasize that this use of the AdS/CFT correspondence is
independent of the use described above because this new CFT is
unrelated to the RS CFT.}. This
requires that the CFT is strongly coupled, in contrast with Starobinsky's
analysis\footnote
{Note that the conformal anomaly is the same at
strong and weak coupling \cite{hensken1} so any differences arising
from strong coupling can only show up when we perturb the system.}.

Our five dimensional (Euclidean) action is the following:
\be \label{eucl}
 S=S_{EH} + S_{GH} + 2S_1 + W_{CFT}.
\ee 
We seek a solution in which two balls of $AdS_5$ are separted by a
spherical domain wall. Inside each ball, the metric can be written
\be
 ds^2 = l^2 (dy^2 + \sinh^2 y d\Omega_d^2),
\ee
with $0 \le y \le y_0$. The domain wall is at $y=y_0$ and has radius
\be
 R = l\sinh y_0.
\ee
The effective tension of the domain wall is given by the Israel
equations as
\be
 \sigma_{eff} = \frac{3}{4\pi G l} \coth y_0.
\ee
The actual tension of the domain wall is
\be
 \sigma = \frac{3}{4\pi G l}.
\ee
We therefore need a contribution to the effective tension from the
CFT. This is provided by the conformal anomaly, which takes the value
\cite{tseytlin,hensken1,hensken2}
\be
 \langle T \rangle = -\frac{3 N^2}{8 \pi^2 R^4},
\ee
This contributes an effective tension 
$-\langle T \rangle /4$. We can now obtain an equation for the radius 
of the domain wall:
\be
\label{eqn:radsol}
 \frac{R^3}{l^3} \sqrt{\frac{R^2}{l^2}+1} = \frac{N^2 G}{8 \pi l^3} +
\frac{R^4}{l^4}.
\ee
It is easy to see that this has a unique positive solution for $R$. We
shall derive this equation directly from the action 
in subsection \ref{subsec:totalaction}.

\medskip

We are particularly interested in how
perturbations of this model would appear to inhabitants of the domain
wall. Thus we are interested in metric perturbations on the sphere
\be
 ds^2 = (R^2 \hat{\gamma}_{ij} + h_{ij}) dx^i dx^j.
\ee
Here $\hat{\gamma}_{ij}$ is the metric on a unit $d$-sphere. We shall
only consider {\it tensor} perturbations, for which $h_{ij}$ is
transverse and traceless with respect to $\hat{\gamma}_{ij}$. In order
to calculate correlators of the metric perturbation, we need to know
the action to second order in the perturbation. The most difficult
part here is obtaining $W_{CFT}$ to second order. This is the subject
of the next subsection.

\subsection{CFT Generating Function}

\label{subsec:CFTgen}

 We want to work out the effect of the perturbation on the CFT on the
sphere. To do this we use AdS/CFT. Introduce a fictional AdS region
that fills in the sphere. Let $\bar{l},\bar{G}$ be the AdS radius and
Newton constant of this region. We emphasize that this region has
nothing to do with the regions of AdS that ``really'' lie inside the
sphere in the RS scenario. This new AdS region is bounded by the
sphere. If we take $\bar{l}$ to zero then the sphere is effectively at
infinity in AdS so we can use AdS/CFT to calculate the generating functional
of the CFT on the sphere. In other words, $\bar{l}$ is acting like a
cut-off in the CFT and taking it to zero corresponds to removing the cut-off.
However the relation
\be
\label{eqn:Ndef2}
 \frac{\bar{l}^3}{\bar{G}} = \frac{2N^2}{\pi},
\ee
implies that if $\bar{l}$ is taken to zero then we must also take
$\bar{G}$ to zero since $N$ is fixed (and large). 

For the unperturbed sphere, the metric in the new AdS region is
\be
 ds^2 = \bar{l}^2 (dy^2 + \sinh^2 y \hat{\gamma}_{ij} dx^i dx^j),
\ee
and the sphere is at $y=y_0$ given by $R=\bar{l}\sinh y_0$. Note
that $y_0 \rightarrow \infty$ as $\bar{l} \rightarrow 0$ since $R$ is
fixed. In order to use AdS/CFT for the perturbed sphere, we need to
know how the perturbation extends into the bulk. This is done by
solving the linearized Einstein equations. It is always possible to
choose a gauge in which the bulk metric perturbation takes the form
\be
 h_{ij}(y,x) dx^i dx^j,
\ee
where $h_{ij}$ is transverse and traceless with respect to the metric
on the spherical spatial sections:
\be
 \hat{\gamma}^{ij}(x) h_{ij}(y,x) = \hat{\nabla}^i h_{ij}(y,x) = 0,
\ee
with $\hat{\nabla}$ denoting the covariant derivative defined by the
metric $\hat{\gamma}_{ij}$. Since we are only dealing with tensor
perturbations, this choice of gauge is consistent with the boundary
sitting at constant $y$. If scalar metric perturbations were
included then we would have to take account of a perturbation in the
position of the boundary. These issues are discussed in detail in
Appendix \ref{app:A}.

The linearized Einstein equations in the bulk are (for any dimension)
\be
\label{eqn:5dperturb}
 \nabla^2 h_{\mu\nu} = -\frac{2}{\bar{l}^2} h_{\mu\nu},
\ee
where $\mu,\nu$ are $d+1$ dimensional indices. It is convenient to
expand the metric perturbation in terms of tensor spherical
harmonics $H^{(p)}_{ij}(x)$. These obey
\be
 \hat{\gamma}^{ij} H^{(p)}_{ij}(x) = \hat{\nabla}^i H^{(p)}_{ij}(x) = 0,
\ee
and they are tensor eigenfunctions of the Laplacian:
\be
 \hat{\nabla}^2 H^{(p)}_{ij} = \left(2-p(p+d-1)\right) H^{(p)}_{ij},
\ee
where $p=2,3,\ldots$. We have suppressed extra labels $k,l,m,\ldots$ on
these harmonics. The harmonics are orthonormal with respect to the
obvious inner product. See Appendix \ref{app:B} and \cite{tensor} 
for more details of their properties. 
The metric perturbation can be written as a sum of separable perturbations
of the form
\be
 h_{ij}(y,x) = f_p (y) H_{ij}^{(p)}(x).
\ee
Substituting this into equation \ref{eqn:5dperturb} gives
\be \label{euclpert}
 f_p''(y) + (d-4) \coth y f_p'(y) - (2(d-2) + (p(p+d-1) + 2(d-3))
\mathrm{cosech}^2 y ) f_p(y) = 0.
\ee
The roots of the indicial equation are $p+2$ and $-p-d+3$, yielding
two linearly independent solutions for each $p$. In order
to compute the generating functional $W_{CFT}$ we
have to calculate the Euclidean action of these solutions.
However, because the latter 
solution goes as $y^{-(p+d-3)}$ at the origin $y=0$ of the instanton, 
the corresponding fluctuation modes have infinite
Euclidean action\footnote{This can be seen by surrounding the origin
by a small sphere $y=\epsilon$ and calculating the surface terms in
the action that arise on this sphere. They are the same as the surface
terms in equations \ref{eqn:perteh} and \ref{eqn:pertgh} below, which
are obviously divergent for the modes in question.}. 
Hence they are suppressed in the path integral.
Therefore, in contrast to other methods \cite{gt,gkr} where one
requires a (rather {\it ad hoc}) prescription for the vacuum state of each
perturbation mode, there is no
need to impose boundary conditions by hand in our approach:
the Euclidean path integral
defines its own boundary conditions, which automatically gives a unique
Green function. The path integral unambiguously specifies the allowed
fluctuation modes as those which vanish at $y=0$. Note that boundary
conditions at the origin in Euclidean space replace the need for boundary
conditions at the horizon in Lorentzian space.
The solution regular at $y=0$ is given by
\be
 f_p (y) = \frac{\sinh^{p+2} y}{\cosh^p y} F(p/2,(p+1)/2, p+(d+1)/2,
\tanh^2 y).
\ee
This solution can also be written in terms of associated Legendre functions:
\be \label{leg}
 f_p (y) \propto (\sinh y)^{(5-d)/2} P^{-(p+(d-1)/2)}_{-(d+1)/2}
(\cosh y) \propto (\sinh y)^{(4-d)/2} Q^{d/2}_{p+(d-2)/2}(\coth y),
\ee
and the latter can be related to Legendre functions if $d/2$ is an
integer, using
\be
 Q^m_{\nu}(z) = (z^2-1)^{m/2} \frac{d^m Q_{\nu}}{dz^m}.
\ee
The full solution for the metric perturbation is
\be
\label{eqn:metricsol}
 h_{ij}(y,x) = \sum_p \frac{f_p(y)}{f_p(y_0)} H^{(p)}_{ij}(x) \int d^dx'
 \sqrt{\hat{\gamma}} h^{kl}(x') H^{(p)}_{kl}(x').
\ee

We have a solution for the metric perturbation throughout the bulk
region. The AdS/CFT correspondence can now be used to give the
generating functional of the CFT on the perturbed sphere:
\be \label{gencft}
 W_{CFT} = S_{EH}+S_{GH}+S_1+S_2+\ldots.
\ee
We shall give the terms on the right hand side for $d=4$. 

The Einstein-Hilbert action with cosmological constant is
\be
 S_{EH} = -\frac{1}{16\pi \bar{G}} \int d^5 x \sqrt{g} \left(R +
\frac{12}{\bar{l}^2} \right),
\ee
and perturbing this gives
\ba
 S_{bulk} &=& -\frac{1}{16 \pi \bar{G}} \int d^5 x \sqrt{g}
\left(-\frac{8}{\bar{l}^2} + \frac{1}{4} h^{\mu\nu} \nabla^2 h_{\mu\nu}
+ \frac{1}{2 \bar{l}^2} h^{\mu\nu} h_{\mu\nu} \right) \nonumber \\ &-& 
\frac{1}{16\pi \bar{G}}
\int d^4 x \sqrt{\gamma} \left( -\frac{1}{2} n^{\mu} h^{\nu\rho}
\nabla_{\nu} h_{\mu\rho} + \frac{3}{4} h_{\nu\rho} n^{\mu}
\nabla_{\mu} h^{\nu\rho} \right),
\ea
where Greek indices are five dimensional and we are raising and
lowering with the unperturbed five dimensional metric. $n = l dy$ is
the unit normal to the boundary and $\nabla$ is the covariant
derivative defined with the unperturbed bulk metric. $\gamma_{ij} =
R^2 \hat{\gamma}_{ij}$ is the unperturbed boundary metric. 
It is important to keep
track of all the boundary terms arising from integration by parts.
Evaluating on shell gives
\be
\label{eqn:perteh}
 S_{EH} = \frac{\bar{l}^3}{2 \pi \bar{G}} \int d^4 x \sqrt{\hat{\gamma}}
\int_0^{y_0} dy 
\sinh^4 y - \frac{\bar{l}^3}{16\pi \bar{G}} \int d^4 x \sqrt{\hat{\gamma}}
\left( \frac{3}{4\bar{l}^4} h^{ij} \partial_y h_{ij} - \frac{\coth
y_0}{\bar{l}^4}  h^{ij} h_{ij} \right).
\ee
where we are now raising and lowering with $\hat{\gamma}_{ij}$. 
The Gibbons-Hawking term is 
\be
\label{eqn:pertgh}
 S_{GH} = -\frac{\bar{l}^3}{2\pi \bar{G}} \int d^4 x
\sqrt{\hat{\gamma}} 
\left(\sinh^3 y_0 \cosh y_0 - \frac{1}{8 \bar{l}^4} h^{ij} \partial_y
 h_{ij}\right). 
\ee
The first counter term is
\ba
\label{eqn:pertct1}
 S_1 &=& \frac{3}{8 \pi \hat{G} \bar{l}} \int d^4 x \sqrt{{\gamma}}
 \nonumber \\      
 &=& \frac{3 \bar{l}^3}{8 \pi \bar{G}} \int d^4x \sqrt{\hat{\gamma}}
 \left(\sinh^4 y_0 - \frac{1}{4\bar{l}^4} h^{ij} h_{ij}\right).
\ea
The second counter term is
\ba
 S_2 &=& \frac{\bar{l}}{32\pi \bar{G}} \int d^4 x \sqrt{\gamma} R \nonumber \\
     &=& \frac{\bar{l}^3}{32 \pi \bar{G}} \int d^4 x \sqrt{\hat{\gamma}} \left(
 12\sinh^2 y_0 - \frac{2}{\bar{l}^4 \sinh^2 y_0}  h^{ij} h_{ij} + \frac{1}{4
 \bar{l}^4 \sinh^2 y_0} h^{ij} \hat{\nabla}^2 h_{ij}\right). 
\ea
Thus with only two counter terms we would have 
\ba
\label{eqn:Wsqrt}
 W_{CFT} = \frac{3 N^2 \Omega_4}{8 \pi^2} \log {\frac{R}{\bar{l}}} &-& 
 \frac{\bar{l}^3}{16\pi
 \bar{G}} \int d^4x \sqrt{\hat{\gamma}} \left(-\frac{1}{4\bar{l}^4} h^{ij}
 \partial_y h_{ij} + \frac{1}{\bar{l}^4} h^{ij} h_{ij} \left(\frac{3}{2} -
 \sqrt{1+ \frac{\bar{l}^2}{R^2}} \right) \right. \nonumber \\ &+& 
 \left. \frac{1}{\bar{l}^2 R^2} h^{ij} h_{ij} -
 \frac{1}{8 \bar{l}^2 R^2} h^{ij} \hat{\nabla}^2 h_{ij} \right).
\ea 
$\Omega_4$ is the area of a unit four-sphere and we have used equation
\ref{eqn:Ndef2}.
The expansion of $\partial_y h_{ij}$ at $y=y_0$ is obtained from
\be
 \partial_y h_{ij} = \sum_p \frac{f_p'(y_0)}{f_p(y_0)} H^{(p)}_{ij}(x)\int d^4
 x'\sqrt{\hat{\gamma}} h^{kl}(x') H^{(p)}_{kl}(x')  
\ee
and
\ba
\label{eqn:fexp}
 \frac{f_p'(y_0)}{f_p(y_0)} &=& 2 + \frac{\bar{l}^2}{2 R^2} (p+1)(p+2) +
p(p+1)(p+2)(p+3) \frac{\bar{l}^4}{4 R^4} \log (\bar{l}/R) +
\frac{\bar{l}^4}{8R^4} \left[ p^4+2p^3 \right. \nonumber \\ {}&-&
\left. 5p^2 -10p -2 -  p(p+1)(p+2)(p+3)
(\psi(1)+\psi(2) -\psi(p/2+2) - \psi(p/2+5/2)) \right] \nonumber \\
{}&+& {\cal O}\left(\frac{\bar{l}^6}{R^6} \log (\bar{l}/R)\right).  
\ea 
The psi function is defined by $\psi(z) = \Gamma'(z)/\Gamma(z)$. 
Substituting into the action we find that the divergences as $\bar{l}
\rightarrow 0$ cancel at
order $R^4/\bar{l}^4$ and $R^2/\bar{l}^2$. 
The term of order $\bar{l}^4/R^4$ in
the above expansion makes a contribution to the finite part of the
action (along with a term from the square root in equation \ref{eqn:Wsqrt}):
\ba
 W_{CFT} &=& \frac{3 N^2 \Omega_4}{8 \pi^2} \log {\frac{R}{\bar{l}}} \\ {} &+&
\frac{N^2}{256 \pi^2 R^4} \sum_p \left(\int d^4x'
 \sqrt{\hat{\gamma}} h^{kl}(x') H^{(p)}_{kl}(x') \right)^2 \left(
2p(p+1)(p+2)(p+3) \log (\bar{l}/R) + \Psi(p)\right), \nonumber
\ea 
where
\ba
 \Psi(p) & = &  p(p+1)(p+2)(p+3) \left[\psi(p/2+5/2) + \psi(p/2+2) -
 \psi(2) - \psi (1)\right]\nonumber\\
& &  + p^4+2p^3-5p^2-10p -6.   
\ea 
To cancel the logarithmic divergences as $\bar{l} \rightarrow 0$,
we have to introduce a length scale
$\rho$ defined by $\bar{l}=\epsilon \rho$ and add a counter term
proportional to $\log \epsilon$ to cancel the divergence as $\epsilon$
tends to zero. The counter term is
\ba
\label{eqn:CT3}
 S_3 &=& -\frac{\bar{l}^3}{64 \pi \bar{G}} \log \epsilon 
 \int d^4 x\sqrt{\gamma}
 \left(\gamma^{ik} \gamma^{jl} R_{ij} R_{kl} - \frac{1}{3} R^2\right)
 \nonumber \\ 
     &=& -\frac{\bar{l}^3}{64 \pi \bar{G}} \log \epsilon \int d^4x
 \sqrt{\hat{\gamma}} \left(-12 + \frac{1}{R^4} \left[
 2h^{ij} h_{ij} - \frac{3}{2} h^{ij} \hat{\nabla}^2 h_{ij} +
 \frac{1}{4} h^{ij} \hat{\nabla}^4 h_{ij} \right]
 \right).
\ea 
This term does indeed cancel the logarithmic divergence, leaving us
with
\ba
\label{eqn:WCFT}
  W_{CFT} &=& \frac{3 N^2 \Omega_4}{8 \pi^2} \log {\frac{R}{\rho}}
 \\ {} &+&
 \frac{N^2}{256 \pi^2 R^4} \sum_p \left(\int d^4x'
 \sqrt{\hat{\gamma}} h^{kl}(x') H^{(p)}_{kl}(x') \right)^2 \left(
 2p(p+1)(p+2)(p+3) \log (\rho/R) + \Psi(p) \right) \nonumber
\ea
Note that varying $W_{CFT}$ twice with respect to $h_{ij}$ yields the
expression for the transverse traceless part of the correlator 
$\langle T_{ij}(x) T_{i'j'}(x')
\rangle$ on a round four sphere. At large $p$, this behaves like $p^4
 \log p$, as expected from the flat space result \cite{gkp}. In fact
this correlator can be determined in closed form 
solely from the trace anomaly and symmetry considerations\footnote{
See \cite{osborn} for a general discussion of such correlators on
maximally symmetric spaces.}. However, we shall be be interested in
calculating cosmologically observable effects, for which our mode
expansion is more useful.

\subsection{The Total Action.}

\label{subsec:totalaction}

Recall that our five dimensional action is 
\be
 S = S_{EH} + S_{GH} + 2S_1 + W_{CFT}.
\ee
In order to calculated correlators of the metric, we need to evaluate
the path integral 
\ba
 Z[{\bf h}] &=& \int_{B_1 \cup B_2} d[\delta{\bf g}] \exp(-S) \\ 
 &=& \exp(-2S_1[{\bf h}_0 + {\bf h}] -
 W_{CFT}[{\bf h}_0 + {\bf h}]) \left(\int_B d[\delta{\bf g}] \exp(-S_{EH}[{\bf g}_0 +
\delta{\bf g}] - S_{GH}[{\bf g}_0 + \delta{\bf g}])\right)^2 \nonumber.
\ea
Here ${\bf g}_0$ and ${\bf h}_0$ refer to the unperturbed background
metrics in the bulk and on the wall respectively and ${\bf h}$ denotes
the metric perturbation on the wall.
Many of the terms required here can be obtained from results in the
previous section by simply replacing $\bar{l}$ and $\bar{G}$ with $l$
and $G$. For example, from equation \ref{eqn:pertct1} we obtain
\be
 S_1[{\bf h}_0 + {\bf h}] = \frac{3 l^3}{8\pi G} \int d^4 x
 \sqrt{\hat{g}} \left( \sinh^4 y_0 - \frac{1}{4l^4} \right),
\ee
where $y_0$ is defined by $R=l\sinh y_0$. The path integral over
$\delta {\bf g}$ is performed by splitting it into a classical and
quantum part:
\be
 \delta{\bf g} = {\bf h} + {\bf h}',
\ee
where the boundary perturbation ${\bf h}$ 
is extended into the bulk using the linearized
Einstein equations and the requirement of finite Euclidean action,
i.e., ${\bf h}$ is given in the bulk by equation \ref{eqn:metricsol}. 
${\bf h}'$ denotes a quantum fluctuation that vanishes at the domain wall. The
gravitational action splits into separate contributions from the classical
and quantum parts:
\be
 S_{EH} + S_{GH} = S_0[{\bf h}] + S'[{\bf h}'],
\ee
where $S_0$ can be read off from equations \ref{eqn:perteh} and
\ref{eqn:pertgh} as
\be
 S_0 = -\frac{3 l^3 \Omega_4}{2\pi G} \int_0^{y_0} dy
 \sinh^2 y_0 \cosh^2 y_0 + \frac{l^3}{16 \pi G}
\int d^4 x \sqrt{\hat{\gamma}} \left(\frac{1}{4l^4} h^{ij} \partial_y h_{ij}
+ \frac{\coth y_0}{l^4} h^{ij} h_{ij} \right),
\ee 
Note that $S'$ cannot be converted to a surface term since ${\bf h}'$ 
does not satisfy the Einstein equations. We shall not need the
explicit form for $S'$ since the path integral over ${\bf h}'$ just
contributes a factor of some determinant $Z_0$ to $Z[{\bf h}]$. We obtain
\be
 Z[{\bf h}] = Z_0 \exp(-2S_0[{\bf h}_0 + {\bf h}] - 2S_1[{\bf h}_0 +
 {\bf h}] - W_{CFT}[{\bf h}_0 + {\bf h}]).
\ee
The exponent is given by
\ba
\label{eqn:pertaction}
 2S_0 + 2S_1 &+& W_{CFT} = -\frac{3l^3\Omega_4}{\pi G} \int_0^{y_0} dy
 \sinh^2 y \cosh^2 y + \frac{3 \Omega_4 R^4}{4 \pi G l} + \frac{3 N^2
 \Omega_4}{8 \pi^2} \log {\frac{R}{\rho}} \nonumber \\ {}
 &+& \frac{1}{l^4} \sum_p \left(\int d^4x'
 \sqrt{\hat{\gamma}} h^{kl}(x') H^{(p)}_{kl}(x') \right)^2
 \left[ \frac{l^3}{32 \pi G} \left( \frac{f_p'(y_0)}{f_p(y_0)} +
 4\coth y_0 -6 \right) \right. \nonumber \\  
 {}&+& \left. 
 \frac{N^2}{256 \pi^2 \sinh^4 y_0}\left( 2p(p+1)(p+2)(p+3) \log (\rho/R) +
 \Psi(p) \right)  \right].  
\ea

\medskip

We have kept the unperturbed action in order to demonstrate how the
conformal anomaly arises: it is simply the coefficient of the $\log
(R/\rho)$ term divided by the area $\Omega_4 R^4$ of the sphere. 
If we set the metric perturbation to zero and vary
$R$ in equation \ref{eqn:pertaction}
(using $R=l\sinh y_0$) then we reproduce equation \ref{eqn:radsol}.

Having calculated $R$, we can now choose a convenient value for the
renormalization scale $\rho$. If we were dealing purely with the CFT
then we could keep $\rho$ arbitrary. However, since the third counter
term (equation \ref{eqn:CT3}) involves the square of the Weyl tensor (the
integrand is proportional to the difference of the Euler density and
the square of the Weyl tensor), we can
expect pathologies to arise if this term is present when we couple the
CFT to gravity. In other words, when coupled to gravity, different
choices of $\rho$ lead to different theories. We shall choose the value
$\rho=R$ so that the third counter term exactly cancels the
divergence in the CFT, with no finite remainder and hence no residual
curvature squared terms in the action.

\medskip

The (Euclidean) graviton correlator can be read off from the action as
\be
\label{eqn:euccor}
 \langle h_{ij}(x) h_{i'j'}(x') \rangle = \frac{128 \pi^2
 R^4}{N^2} \sum_{p=2}^{\infty}
 W^{(p)}_{ij i'j'}(x,x') F(p,y_0)^{-1}
\ee
where we have eliminated $l^3/G$ using equation \ref{eqn:radsol}. 
The function $F(p,y_0)$ is given by
\be 
 F(p,y_0) = e^{y_0} \sinh y_0 \left(\frac{f_p'(y_0)}{f_p(y_0)} + 4\coth y_0 -6
 \right) + \Psi(p),
\ee
and the bitensor $W^{(p)}_{iji'j'}(x,x')$ is defined as
\be
 W^{(p)}_{iji'j'}(x,x') = \sum_{k,l,m,\ldots} H^{(p)}_{ij}(x)
 H^{(p)}_{i'j'}(x'),
\ee
with the sum running over all the suppressed labels $k,l,m,\ldots$ of the
tensor harmonics. 

The appearance of $N^2$ in the denominator in equation
\ref{eqn:euccor} suggests that the CFT suppresses metric perturbations
on all scales. This is misleading because $R$ also depends on $N$. 
The function $F(p,y_0)$ has the following limiting forms for large
and small radius:
\be
\label{eqn:largeR}
 \lim_{y_0\rightarrow \infty} F(p,y_0) = \Psi(p) + p^2 + 3p +6,
\ee
\be
\label{eqn:smallR}
 \lim_{y_0 \rightarrow 0} F(p,y_0) = \Psi(p) + p + 6.
\ee
$F(p,y_0)$ has  poles at $p = -4,-5,-6,\ldots$
with zeros between each pair of negative integers starting at $-3,-4$. 
When we analytically continue to Lorentzian signature, we shall be
particularly interested in zeros lying in the range $p \ge -3/2$.
There is one such zero exactly at $p=0$, another near $p=0$ and a third near
$p=-3/2$. For large radius, these extra zeros are at $p \approx -0.054$
and $p\approx -1.48$ while for small radius they are at $p \approx
0.094$ and $p \approx -1.60$. For intermediate radius they lie between
these values, with the zeros crossing through $-3/2$ and $0$ at $y_0
\approx 0.632$ and $y_0 \approx 1.32$ respectively. 

\subsection{Comparison With Four Dimensional Gravity.}

We discussed in section \ref{sec:RSCFT} how the RS scenario reprodues
the predictions of four dimensional gravity when the effects of matter
on the domain wall dominates the effects of the RS CFT. In our case we
have a CFT on the domain wall. This has action proportional to
$N^2$. The RS CFT is a similar CFT (but with a cut-off) and therefore
has action proportional to $N_{RS}^2$. Hence we can neglect it when $N
\gg N_{RS}$. The logarithmic counterterm is also proportional to
$N_{RS}^2$ and therefore also negligible. We therefore expect the
predictions of four dimensional gravity to be recovered when $N \gg
N_{RS}$. We shall now demonstrate this explicitly.

First consider the radius $R$ of the domain wall given by equation
\ref{eqn:radsol}. It is convenient to write this in terms of the 
rank $N_{RS}$ of the RS CFT (given by $l^3/G=2N_{RS}^2/\pi$)
\be
 \frac{R^3}{l^3} \sqrt{\frac{R^2}{l^2}+1} = \frac{N^2}{16 N_{RS}^2} +
\frac{R^4}{l^4}.
\ee
If we assume $N \gg N_{RS} \gg 1$ then the solution is
\be
 \frac{R}{l} = \frac{N}{2\sqrt{2} N_{RS}} \left[1+\frac{N_{RS}^2}{N^2}
+ {\cal O}(N_{RS}^4/N^4)\right].
\ee
Note that this implies $R \gg l$, i.e., the domain wall is large
compared with the anti-de Sitter length scale.

Now let's turn to a four dimensional description in which we are considering
a four sphere with no interior. The only matter present is the CFT. 
The metric is simply
\be
 ds^2 = R_4^2 \hat{\gamma}_{ij} dx^i dx^j,
\ee
where $R_4$ remains to be determined. The action is the four
dimensional Einstein-Hilbert action (without cosmological constant) 
together with $W_{CFT}$.
There is no Gibbons-Hawking term because there is no boundary. 
Without a metric perturbation, the action is simply
\be
 S = -\frac{1}{16 \pi G_4} \int d^4 x \sqrt{{\gamma}}R +  W_{CFT} 
 = -\frac{3 \Omega_4 R_4^2}{4 \pi G_4} + \frac{3
 N^2 \Omega_4}{8 \pi^2} \log {\frac{R_4}{\rho}}.
\ee
where $G_4$ is the four dimensional Newton constant. 
We want to calculate the value of
$R_4$ so we can't choose $\rho = R_4$ yet. Varying $R_4$ gives
\be
 R_4^2 = \frac{N^2 G_4}{4 \pi},
\ee
and $N$ is large hence $R_4$ is much greater than the four dimensional
Planck length. Substituting $G_4 = G_5/l$, this reproduces the leading
order value for $R$ found above from the five dimensional calculation.

We can now go further and include the metric perturbation. The
perturbed four dimensional Einstein-Hilbert action is
\be
\label{eqn:pertbulk}
 S^{(4)}_{EH} = -\frac{1}{16 \pi G_4} \int d^4 x \sqrt{\hat{\gamma}} \left(
 12 R_4^2  - \frac{2}{R_4^2}  h^{ij} h_{ij} + \frac{1}{4 R_4^2} h^{ij} 
 \hat{\nabla}^2 h_{ij}\right).
\ee
Adding the perturbed CFT gives
\ba
\label{eqn:4daction}
 S &=& -\frac{3 N^2 \Omega_4}{16\pi^2} + \frac{3 N^2 \Omega_4}{8 \pi^2}
 \log {\frac{R_4}{\rho}} + \sum_p
 \left(\int d^4x' \sqrt{\hat{\gamma}} h^{kl}(x') H^{(p)}_{kl}(x') \right)^2 
 \left[ \frac{1}{64 \pi G_4 R_4^2} (p^2+3p+6) \right. \nonumber \\  
 {}&+& \left. 
 \frac{N^2}{256 \pi^2 R_4^4}\left( 2p(p+1)(p+2)(p+3) \log (\rho/R_4) +
 \Psi(p) \right)  \right].
\ea
Setting $\rho=R_4$, we find that the graviton correlator for a four
dimensional universe containing the CFT is
\be
\label{eqn:4dcorr}
 \langle h_{ij}(x) h_{i'j'}(x') \rangle = 8 N^2 G_4^2
 \sum_{p=2}^{\infty} W_{iji'j'}^{(p)}(x,x')
 \left[p^2 + 3p +6 + \Psi(p) \right]^{-1}.
\ee
This can be compared with the expression obtained from the five
dimensional calculation, which can be written
\ba
 \langle h_{ij}(x) h_{i'j'}(x') \rangle &=& \frac{8 N^2 G^2}{l^2}
 \left[1+{\cal O}(N_{RS}^2/N^2)\right]\sum_{p=2}^{\infty}
 W_{iji'j'}^{(p)}(x,x')  \left[p^2 +
 3p +6 + \Psi(p) \right. \\ {} &+& \left.
  4p(p+1)(p+2)(p+3)(N_{RS}^2/N^2) \log
 (N_{RS}/N) + {\cal O}(N_{RS}^2/N^2)\right]^{-1}. \nonumber
\ea
We have expanded in terms of
\be
 \frac{N_{RS}^2}{N^2} = \frac{\pi l^3}{2N^2 G}.
\ee
The four and five dimensional expressions clearly agree (for $G_4 =
G/l$) when $N \gg
N_{RS}$, i.e., $R \gg l$. There are corrections of order
$(N_{RS}^2/N^2) \log (N_{RS}/N)$ coming from the RS CFT and the
logarithmic counter term. In fact, these corrections can be absorbed
into the renormalization of the CFT on the domain wall if, instead
of choosing $\rho = R$, we choose
\be
 \rho = R \left(1 - \frac{2 N_{RS}^2}{N^2} \log (N_{RS}/N)\right).
\ee
The corrections to the four dimensional expression are then of order
$N_{RS}^2/N^2$. We shall not give these correction terms explicitly
although they are easily obtained from the exact result \ref{eqn:euccor}. 

\subsection{Lorentzian Correlator.}

\label{subsec:continue}

In this subsection we shall show how the Euclidean correlator
calculated above is analytically continued to give a correlator
for Lorentzian signature. We have put many of the details
in Appendix \ref{app:B} but the analysis is still rather technical so the
reader may wish to skip to the final result, which is given in equation
\ref{lcor}. The techniques used here were developed in
\cite{Gratton,hertog,hht}. 

\medskip

Let us first introduce a new label $p'=i(p+3/2)$, so that on the four sphere
\be
 \hat{\nabla}^2 H_{ij}^{(p')} = \lambda_{p'} H_{ij}^{(p')},
\ee
where $p'=7i/2,9i/2,...$ and
\be
 \lambda_{p'} = (p'^2 +17/4).
\ee
Recall that there are extra labels on the tensor harmonics
that we have suppressed. The set of rank-two 
tensor eigenmodes on $S^4$ forms a representation
of the symmetry group of the manifold. Hence the sum (equation \ref{bit})
of the degenerate eigenfunctions with eigenvalue
$\lambda_{p'}$ defines a maximally symmetric bitensor 
$W^{\ ij}_{(p')\ i'j'}(\mu (\Omega,\Omega'))$,
where $\mu (\Omega,\Omega')$ is the distance along the shortest geodesic
between the points with polar angles $\Omega$ and $\Omega'$.
The expression of the bitensor in terms of a set of fundamental bitensors
with $\mu$-dependent coefficient functions 
together with the relation between the bitensors on $S^4$ and 
Lorentzian de Sitter space are obtained in Appendix \ref{app:B}.

The motivation for the unusual labelling is that, as demonstrated in
Appendix \ref{app:B}, in terms of the label $p'$ the bitensor on
$S^4$ has exactly the same formal expression as the corresponding
bitensor on Lorentzian de Sitter space.
This property will enable us to analytically continue
the Euclidean correlator into the Lorentzian region without
Fourier decomposing it. In other words, 
instead of imposing by hand a prescription for the
vacuum state of the graviton on each mode separately and 
propagating the individual modes into the Lorentzian region, we
compute the two-point tensor correlator in real space, directly
from the no boundary path integral.
Since the path integral unambiguously specifies the allowed
fluctuation modes as those which vanish at the origin of the instanton
(see discussion in subsection \ref{subsec:CFTgen}),
this automatically gives a unique Euclidean correlator.  
The technical advantage of our method is that
dealing directly with the real space correlator
makes the derivation independent of the gauge
ambiguities involved in the mode decomposition \cite{hht}.

We begin by continuing the graviton correlator (equation
\ref{eqn:euccor}) obtained
via the five dimensional calculation. The analytic continuation of 
the correlator for four dimensional gravity
(equation \ref{eqn:4dcorr}) is completely analogous. 
In terms of the new label $p'$, 
the Euclidean correlator \ref{eqn:euccor} between two points on the wall
is given by
\be \label{cor1}
\langle  h_{ij}(\Omega) h_{i'j'}(\Omega ')\rangle = 
\frac{128 \pi^2 R^4}{N^2} 
\sum_{p'=7i/2}^{i\infty} W_{\ iji'j'}^{(p')}(\mu) G(p',y_{0})^{-1}
\ee
where 
\ba
G(p',y_{0}) & = & F(-ip'-3/2,y_{0}) \\
& = & e^{y_{0}}\sinh y_{0} \left( \frac{g_{p'}' (y_{0})}{g_{p'} (y_{0})}
+4 \coth y_{0} -6 \right)  
+ \left( p'^4 -4ip'^3 +p'^2/2 -5ip' -63/16 \right.
\nonumber\\
& + & \left. (p'^2+1/4)(p'^2+9/4)
[\psi(-ip'/2 +5/4) + \psi(-ip'/2+7/4) -\psi (1) -\psi (2)] \right). \nonumber
\ea
with
$g_{p'}(y) = Q^{2}_{-ip'-1/2} (\coth y)$, which follows from eq. \ref{leg}.
The function $G(p',y_{0})$ is real and positive for all values of
$p'$ in the sum and for arbitrary $y_{0} \ge 0$. 

We have the Euclidean correlator defined as an infinite sum. However,
the eigenspace of the Laplacian on de Sitter space suggests that
the Lorentzian propagator is most naturally expressed as an integral over real
$p'$. We must therefore first analytically continue our result from
imaginary to real $p'$. 
The coefficient $G(p',y_{0})^{-1}$ of the
bitensor is analytic in the upper half complex $p'$-plane,
apart from three simple poles on the imaginary axis.
One of them is always at $p' =3i/2$, regardless of the 
radius of the sphere. Let the position of the remaining two poles 
be written $p_{k}' = i \Lambda_{k}(y_{0})$.
If we take the radius of the domain wall to be
large compared with the AdS scale 
(which is necessary for corrections to four dimensional Einstein
gravity to be small)
then\footnote{If 
we decrease the radius of the
domain wall, then
the poles move away from each other. Their behaviour follows from the
discussion below equations \ref{eqn:largeR} and \ref{eqn:smallR}.
For $y_{0} \leq 0.632$, 
$\Lambda_{1}$ becomes slightly smaller than zero while for
$y_{0} \leq 1.32$, $\Lambda_{2}$ becomes slightly greater than $3/2$.}
$0 < \Lambda_{k} \leq 3/2$, with
$\Lambda_1 \sim 0$ and $\Lambda_2 \sim 3/2$.
Since $G(p',y_{0})$ is real on the imaginary $p'$-axis, the residues
at these poles are purely imaginary.
In order to extend the correlator into the
complex $p'$-plane, we must also understand the continuation of
the bitensor itself. As shown in Appendix \ref{app:B}, 
the condition of regularity at
opposite points on the four sphere
imposed by the completeness relation (equation \ref{com})
is sufficient to uniquely specify the analytic 
continuation of $W_{\ iji'j'}^{(p')}(\mu)$ 
into the complex $p'$-plane. The extended bitensor is defined by equations
\ref{bitensor}, \ref{app1} and \ref{relat}.

Now we are able to write the sum in equation \ref{cor1} as an integral
along a contour ${\cal C}_1$ encircling the points
$p'=7i/2,9i/2,..ni/2$, where $n$ tends to infinity.
This yields
\be \label{cor2a}
\langle  h_{ij}(\Omega) h_{i'j'}(\Omega ')\rangle
 = \frac{-i64 \pi^2 R^4}{N^2}
\int_{{\cal {C}}_1} dp' \tanh p'\pi
W_{\ iji'j'}^{(p')}(\mu) G(p',y_{0})^{-1}.
\ee

Since we know the analytic properties of the integrand in the
upper half complex $p'$-plane,
we can distort the contour for the $p'$ integral to run along
the real axis. At large imaginary $p'$ the integrand decays and the 
contribution vanishes in the large $n$ limit.
However as we deform the contour 
towards the real axis, we encounter 
three extra poles in the $\cosh p'\pi$ factor, the pole at $p'=3i/2$
becoming a double pole due to the simple zero of $G(p',y_{0})$. 
In addition, we have to take in account
the two poles of $G(p',y_{0})^{-1}$ at $p'=i\Lambda_{k}$.

For the $p'=5i/2$ pole, 
it follows from the normalization of the tensor 
harmonics that $W_{\ iji'j'}^{(5i/2)} =0$. Indirectly, this is a consequence of
the fact that spin-2 perturbations do not have a dipole or monopole 
component.
The meaning of the remaining two poles of the $\tanh p' \pi$
factor has been extensively discussed 
in \cite{hht},  where the
continuation is described of the two-point tensor 
fluctuation correlator
from a four dimensional $O(5)$ instanton into open de Sitter space.
They represent non-physical contributions to the graviton propagator,
arising from the different nature of tensor harmonics on
$S^4$ and on Lorentzian 
de Sitter space. In fact, a degeneracy appears between 
$p'_{t}=3i/2$ and $p'_{t}=i/2$ 
tensor harmonics and respectively $p'_{v}=
5i/2$ vector harmonics and $p'_{s}=5i/2$ 
scalar harmonics on $S^4$. More precisely, the tensor harmonics that
constitute the bitensors $W^{\ iji'j'}_{(3i/2)}$ and
$W^{\ iji'j'}_{(i/2)}$ can be constructed from a vector (scalar) quantity.
Consequently, the contribution to the correlator from the
former pole is pure gauge, while the latter eigenmode 
should really be treated as a scalar 
perturbation, using the perturbed scalar action. 
Henceforth we shall exclude them from the tensor spectrum.
This leaves us with the poles of $G(p',y_{0})$ at $p'=i\Lambda_{k}$.
If we deform the contour towards the real axis, we must compensate for
them by subtracting their residues from the integral.
We will see that these residues correspond to 
discrete ``supercurvature'' modes in the Lorentzian tensor correlator.

The contribution from the closing of the contour in the
upper half $p'$-plane vanishes.
Hence our final result for the Euclidean correlator reads
\ba \label{cor2}
\langle  h_{ij}(\Omega) h_{i'j'}(\Omega ')\rangle
& =  &  \frac{-i64 \pi^2 R^4}{N^2} \left[
\int_{-\infty}^{+\infty} dp' \tanh p'\pi
W_{\ iji'j'}^{(p')}(\mu)G(p',y_{0})^{-1} \right.\nonumber\\
& & \left. +2\pi \sum_{k=1}^{2} \tan \Lambda_{k} \pi
W_{\ iji'j'}^{(i\Lambda_{k})}(\mu){\bf Res }(
G(p',y_{0})^{-1};i\Lambda_{k}) \right].
\ea

The analytic continuation from a four sphere
into Lorentzian closed de Sitter space is given by setting
the polar angle $\Omega =\pi /2 -it$. Without loss of
generality we may take $\mu =\Omega$, and $\mu$ then 
continues to $\pi /2 -it$. We then obtain the correlator in
de Sitter space where one point has been chosen as the origin of the 
time coordinate.

The continuation of the bitensor $W_{\ iji'j'}^{(p')} (\mu)$
is given in Appendix \ref{app:B}.
An extra subtlety arises if one wants to
identify the continued bitensor with the usual sum of
tensor harmonics on de Sitter space.
It turns out that in order to do so,
one must extract a factor $ie^{p\pi }/\sinh p' \pi $ from its coefficient 
functions\footnote{The underlying reason
is that there exist two independent bitensors of the form defined by
equations \ref{bitensor} and \ref{app1}. Under the integral
in the Lorentzian
correlator, they are related by the factor $ie^{p\pi }/\sinh p' \pi $. 
It follows from the continuation of the completeness
relation (equation \ref{com}) 
that the sum of degenerate tensor harmonics on de Sitter space equals
the second independent bitensor, rather then the  
bitensor that we obtain by continuation from $S^4$.
Therefore, in order to express the Lorentzian two-point
tensor correlator in terms of tensor harmonics, we must extract
this factor from the bitensor. 
We refer the interested reader to the Appendix for the
details.}.
We denote the final form of the bitensor by
$W_{\ iji'j'}^{L(p')}(\mu (x,x') )$, which is
defined in the Appendix, equations \ref{bitensor},
\ref{app1} and \ref{app3}.  

The extra factor $ie^{p\pi }/\sinh p' \pi $ combines with the factor
$-i \tanh p' \pi $ in the integrand to $e^{p' \pi }/\cosh p' \pi $.
Furthermore, since $ G(-p',y_{0})= \bar G (p',y_{0})$, 
we can rewrite the correlator
as an integral from $0$ to $\infty$. We finally obtain the Lorentzian
tensor Feynman (time-ordered) correlator,
\ba \label{lcor}
\langle  h_{ij}(x) h_{i'j'}(x')\rangle
 & = & \frac{128 \pi^2 R^4}{N^2} \left[\int_{0}^{+\infty} dp'
 \tanh p'\pi W_{\ iji'j'}^{L(p')}(\mu)\Re ( G(p',y_{0})^{-1}) \right.
\nonumber\\ 
& & \qquad \qquad \qquad \left.
+\pi \sum_{k=1}^{2}
\tan \Lambda_{k} \pi  W_{\ iji'j'}^{L(i\Lambda_{k})}(\mu) {\bf Res}
(G(p',y_{0})^{-1};i\Lambda_{k})\right]\nonumber\\
& & +i\frac{128 \pi^2 R^4}{N^2} \left[\int_{0}^{+\infty} dp'
W_{\ iji'j'}^{L(p')}(\mu)\Re ( G(p',y_{0})^{-1}) \right. \nonumber\\
& & \left. \qquad \qquad \qquad
-\pi \sum_{k=1}^{2}
W_{\ iji'j'}^{L(i\Lambda_{k})}(\mu) {\bf Res}
(G(p',y_{0})^{-1};i\Lambda_{k})\right].
\ea
In this integral the bitensor  $W_{\ iji'j'}^{L(p')}(\mu (x,x') )$ may
be written as
the sum of the degenerate rank-two tensor harmonics on closed de Sitter space
with eigenvalue  $\lambda_{p'}= ({p'}^2 +17/4)$ of the Laplacian.
Note that the normalization factor $\tilde Q_{p'}=
p'(4p'^2 +25)/48 \pi^2$ of the bitensor is imaginary at $p'
=i\Lambda_k$ and the residues of $G^{-1}$ are also imaginary, 
so the quantities in square brackets are all real. 
Both integrands in equation \ref{lcor}
vanish as $p' \rightarrow 0$, so the correlator is
well-behaved in the infrared. 

For cosmological applications, one is usually interested in the
expectation
of some quantity squared, like the microwave background multipole
moments. For this purpose, all that matters is the symmetrized
correlator, which is just the real part of the
Feynman correlator. 

Gravitational waves provide an extra source of time-dependence
in the background in which the cosmic microwave background photons
propagate.
In particular, the contribution of gravitational waves to the
CMB anisotropy is given by the integral in the Sachs-Wolfe formula,
which is basically the integral 
along the photon trajectory of the time derivative
of the tensor perturbation.
Hence the resulting microwave multipole moments ${\cal
C}_{l}$ can be directly determined 
from the graviton correlator.

We can therefore understand the effect of the strongly coupled CFT on the
microwave 
fluctuation spectrum by comparing our result \ref{lcor} with the transverse
traceless part of the graviton propagator in four-dimensional
de Sitter spacetime \cite{turyn}.
On the four-sphere, this is easily obtained by varying the
Einstein-Hilbert action with a cosmological constant. 
In terms of the bitensor, this yields
\ba \label{scor1}
\langle  h_{ij}(\Omega) h_{i'j'}(\Omega ')\rangle & = & 
32 \pi G_4 R^2 \sum_{p'=7i/2}^{i\infty} 
\frac{W_{\ iji'j'}^{(p')}(\mu (\Omega,\Omega'))}
{\lambda_{p'}-2},
\ea
which continues to
\ba \label{slcor}
\langle  h_{ij}(x) h_{i'j'}(x')\rangle
 & = & 32 \pi G_4 R^2 \int_{0}^{+\infty} \frac{dp'}{\lambda_{p'}-2}
W_{\ iji'j'}^{L(p')}(\mu(x,x')).
\ea
This can be compared with equation \ref{lcor}.
Note that (apart from the pole at $p'=3i/2$ corresponding to the gauge
mode mentioned before) there are no supercurvature modes.
We defer a detailed discussion of the effect of the CFT on the tensor
perturbation spectrum in de Sitter space to the next section.

\sect{Conclusion}

\label{sec:conclude}

We have studied a Randall-Sundrum cosmological scenario consisting of
a domain wall in anti-de Sitter space with a large
$N$ conformal field theory living on the wall.
The confomal anomaly of the CFT provides an effective tension which
leads to a de Sitter geometry for the domain wall.
We have computed the spectrum of quantum mechanical vacuum 
fluctuations of the graviton field on the domain wall,
according to Euclidean no boundary initial conditions. 
The Euclidean path integral unambiguously specifies
the tensor correlator with no additional assumptions.
This is the first calculation of quantum fluctuations for RS
cosmology. 

In the usual inflationary models, one considers the classical action
for a single scalar field.
In that context, it is consistent to neglect quantum matter loops, on
the grounds that they are small.
On the other hand, in this paper we have studied a strongly coupled large $N$
CFT living on the domain wall, for which quantum loops of matter are
important.
By using the AdS/CFT correspondence, we have 
performed a fully quantum mechanical treatment of this CFT.
The most notable effect of the large $N$ CFT on the tensor spectrum 
is that it suppresses
small scale fluctuations on the microwave sky. 
It can be seen from equation \ref{lcor}
that the CFT yields a $(p'^{4}\ln p')^{-1}$ behaviour 
 for the graviton propagator at large $p'$
(in agreement with the
flat space results of \cite{tom}),
instead of the usual $p'^{-2}$ falloff  (equation \ref{slcor}).
In other words, quantum loops of the 
CFT give spacetime a rigidity that strongly suppresses metric
fluctuations on small scales.
Note that this is true independently of how the de Sitter geometry
arises, i.e. it is also true for four dimensional Einstein gravity.
In addition, the coupling of the CFT to tensor perturbatons
gives rise to two additional discrete modes in the tensor spectrum.
Although this is a novel feature in the context of inflationary tensor
perturbations, it is not surprising.
In conventional open inflationary scenarios for instance,
the coupling of scalar field fluctuations with scalar
metric perturbations introduces a supercurvature mode with an
eigenvalue of the Laplacian
close to the discrete de Sitter gauge mode \cite{yam,Gratton}.
The former discrete mode at $p'=i\Lambda_{1} \sim 3i/2$ in equation \ref{lcor}
is nothing else than the analogue of
this well known supercurvature mode in the scalar fluctuation
spectrum.
The second mode has an eigenvalue $p'=i\Lambda_{2} \sim 0$.
Its interpretation is less clear, but it 
is clearly an effect of the matter on the domain wall. 
However it hardly
contributes to the correlator because $\tan \Lambda_{2} \pi$
is very small.

The effect of the CFT on large scales is more difficult to
quantify because of the complicated $p'$-dependence of
the tensor correlator (equation \ref{lcor}) in the low-$p'$ regime.
Generally speaking, however, long-wavelength tensor correlations
in closed (or open) models for inflation
are very sensitive to the details of the underlying
theory, as well as to the boundary conditions at the instanton. 
Since tensor fluctuations do give a substantial contribution to the 
large scale CMB anisotropies, this may provide an additional
way to observationally distinguish different inflationary scenarios \cite{ght}.

Most matter fields can be expected to behave like a CFT at small
scales. Furthermore, fundamental theories such as string theory
predict the existence of a large number of matter
fields. Therefore, our results based on a quantum treatment of a large
$N$ CFT may be accurate at small scales for any matter. 
If this is the case then our result shows that tensor 
perturbations at small angular scales are much smaller than
predicted by calculations that neglect quantum effects of matter
fields. 

\medskip

\centerline{\bf Acknowledgments}

\noindent It is a pleasure to thank Steven Gratton, 
Hugh Osborn and Neil Turok for useful discussions.

\appendix

\renewcommand{\theequation}{\Alph{section}.\arabic{equation}}

\section{Choice of Gauge}

\label{app:A}

\setcounter{equation}{0}

This appendix demonstrates how a metric
perturbation on the boundary of a ball of AdS is decomposed into
vector, scalar and tensor components.

\medskip

Consider a ball of perturbed AdS with a spherical boundary. 
Let $\bar{l}$ be the AdS length scale. Gaussian normal coordinates are
introduced by defining $\bar{l} y$ to be the geodesic distance of a
point from the origin. The surfaces of constant $y$ are spheres on
which we introduce coordinates $x^i$. In these coordinates
the metric takes the form
\be
 ds^2 = \bar{l}^2(dy^2 + \sinh^2 y \hat{\gamma}_{ij}(x)
dx^i dx^j) + h_{ij}(y,x) dx^i dx^j.
\ee
The ball of AdS has been perturbed, so the boundary will be at a
position $y=y_0 + \xi(x)$. 

Let the induced metric perturbation on the boundary be
$\hat{h}_{ij} (x)$. This can be decomposed into scalar,
vector and tensor perturbations with respect to the round metric on 
the sphere \cite{stewart}:
\be
\label{eqn:decomp}
 \hat{h}_{ij}(x) = \hat{\theta}_{ij} +
2\hat{\nabla}_{(i} \hat{\chi}_{j)} + \hat{\nabla}_i \hat{\nabla}_j
\hat\phi + \hat{\gamma}_{ij} \hat{\psi},
\ee
where we use hats to denote quantities defined on the
sphere (i.e. quantities that depend only on $x$). 
$\hat{\theta}_{ij}$ is a transverse traceless tensor on the
sphere and $\hat{\chi}_i$ is a transverse vector on the
sphere. $\hat{\phi}$ and $\hat{\psi}$ are scalars on the sphere.
$\hat{\chi}_i$ and $\hat{\phi}$ can be gauged away by infinitesimal
coordinate transformations on the sphere of the form $x^i =
\tilde{x}^i - \eta^i(\tilde{x})-\partial^i \eta(\tilde{x})$ where
$\eta^i$ is transverse. Therefore we shall assume that $\hat{\chi}$ and
$\hat{\phi}$ vanish. Note that it is not possible to gauge away
$\hat{\psi}$ or $\xi$. 
This paper only deals with tensor perturbations so we shall
assume that the scalars $\hat{\psi}$ and $\xi$ are vanishing. The induced
metric perturbation is then transverse and traceless and can be
extended into the bulk as described in section 
\ref{sec:CFT}. The scalars will be discussed in our next paper.

\section{Maximally Symmetric Bitensors.}

\label{app:B}

\setcounter{equation}{0}

A maximally symmetric bitensor $T$ is one for which $\sigma^{*}T=0$
for any isometry $\sigma$ of the maximally symmetric manifold.
Any maximally symmetric bitensor may be expanded in terms of a complete set of
fundamental maximally symmetric bitensors with the correct index symmetries.
For instance
\begin{eqnarray}\label{maxi}
T_{iji'j'} &  = & 
t_1(\mu) g_{ij}^{\ }g_{i'j'}^{\ }+
t_2(\mu)n_{(i}^{\ }g_{j)(i'}^{\ }n_{j')}^{\ }+t_3(\mu)
\left[ g_{ii'}^{\ }g_{jj'}^{\ }+g_{ji'}^{\ }g_{ij'}^{\ }
\right] \nonumber\\
& & + t_4(\mu)n_{i}^{\ }n_{j}^{\ }n_{i'}^{\ }n_{j'}^{\ }
+t_5(\mu)\left[g_{ij}^{\ }n_{i'}^{\ }n_{j'}^{\ }+n_{i}^{\ }n_{j}^{\ }
g_{i'j'}^{\ }\right].
\end{eqnarray}
The coefficient functions $t_{j}(\mu)$ depend only on the 
distance $\mu(\Omega,\Omega')$ along the shortest geodesic
from the point $\Omega$ to the point $\Omega'$.
$n_{i'}^{\ }(\Omega,
\Omega ')$ and $n_{i}^{\ }(\Omega, \Omega ')$ are
unit tangent vectors to the geodesics joining $\Omega$ and $\Omega'$ and
$g_{ij'}(\Omega, \Omega ')$ is the parallel propagator along the 
geodesic, i.e., $V^{i}g_{i}^{j'}$ is the vector at $\Omega'$ obtained by
parallel transport of $V^{i}$ along the geodesic from $\Omega$ to $\Omega'$
\cite{Jacob}.

The set of tensor eigenmodes on $S^4$ (or on de Sitter space)
forms a representation of the symmetry group of the
manifold. It follows in particular that 
their sum over the parity states  ${\cal P}=\{e,o\}$ and the
quantum numbers $k,l$ and $m$ on the three
sphere defines a maximally symmetric 
bitensor on $S^4$ (or dS space) \cite{Jacob}:
\begin{equation}\label{bit}
W^{\ ij}_{(p')\ i'j'}(\mu) =
\sum_{{\cal P}klm} q^{(p')ij}_{{\cal P}klm}(\Omega ) 
q^{(p'){\cal P}klm}_{i'j'}(\Omega ')^{*}.
\end{equation}
On $S^4$ the label $p'$ takes the value $7i/2,9i/2,..$. 
It is related to a real label $p$ by $p'=i(p+3/2)$. The ranges of the
other labels are then $0 \le k \le p$, $0 \le l \le k$ and $-l \le m \le
l$. On de Sitter space 
there is a continuum of eigenvalues $p' \in [ 0, \infty)$.
We will assume from now on that the eigenmodes are normalized by the condition
\begin{equation}\label{norm}
\int \sqrt{\gamma} d^{4} \Omega q^{(p')ij}_{{\cal P}klm}
q_{{\cal P}'k'l'm'ij}^{(p'')*} = \delta^{p'p''}\delta_{{\cal P}{\cal P}'}
\delta_{ll'}\delta_{mm'}
\end{equation}
The completeness relation on the four sphere may then be written as
\begin{equation}\label{com}
\gamma^{-{1\over 2}}
\delta^{ij}_{\ \ i'j'}(\Omega - \Omega ') = \sum_{p'=7i/2}^{+i\infty} 
W^{\ ij}_{(p')\ i'j'}(\mu(\Omega,\Omega')).
\end{equation}
Explicit formulae for the components of these tensors may be found
in \cite{tensor}.
In this Appendix we will determine $W_{\ iji'j'}^{(p')}(\mu )$ 
simultaneously on the four sphere and de Sitter space.
The construction of the analogous bitensor on $S^3$ and $H^3$ is given
in \cite{Allen} and their relation is described in \cite{hht}.

The bitensor $W^{\ ij}_{(p')\ i'j'}(\mu)$ has some additional 
properties arising from its construction in terms of the transverse and 
traceless tensor harmonics $q_{ij}^{(p){\cal P}klm}$.
The tracelessness of $W^{(p')}_{iji'j'}$
allows one to eliminate two of the coefficient functions in
equation \ref{maxi}.
It may then be written as
\begin{eqnarray}\label{bitensor}
W^{(p')}_{iji'j'}(\mu) & = & 
w^{(p')}_1\left[ g_{ij}^{\ } -4n_{i}^{\ }n_{j}^{\ }\right]
\left[g_{i'j'}^{\ } -4n_{i'}^{\ }n_{j'}^{\ }\right]
+ w_2^{(p')}\left[4 n_{(i}^{\ }g_{j)(i'}^{\ }n_{j')}^{\ }
+4n_{i}^{\ }n_{j}^{\ }n_{i'}^{\ }n_{j'}^{\ }
\right]\cr
& & 
+w_3^{(p')}\left[ g_{ii'}^{\ }g_{jj'}^{\ }+g_{ji'}^{\ }g_{ij'}^{\ }
 -2n_{i}^{\ }g_{i'j'}^{\ }n_{j}^{\ } -2n_{i'}^{\ }g_{ij}^{\ }n_{j'}^{\ }
+8n_{i}^{\ }n_{j}^{\ }n_{i'}^{\ }n_{j'}^{\ }\right]
\end{eqnarray}
 This expression is traceless on either the index pair
$ij$ or $i'j'$.
The requirement that the bitensor be transverse
$\nabla^{i} W_{iji'j'}^{(p')}=0$ and the
eigenvalue condition $(\nabla^2 - \lambda_{p'})W^{\ iji'j'}_{(p')}=0$
impose additional constraints on the remaining coefficient functions
$w_{j}^{(p')}(\mu)$. To solve these constraint equations it is convenient to
introduce the new variables on $S^4$ (in de Sitter space, $\mu$ is replaced by
$\pi /2 -i\tilde \mu$)
\begin{equation}\label{bet}
\left\{
\begin{array}{lll}
\alpha(\mu) & = &  w_1^{(p)}(\mu) + 
\frac{2}{3}w_3^{(p)}(\mu)\\
\beta(\mu)&  = & \frac{8}{(\lambda_{p} +8)\sin \mu}\frac{d\alpha(\mu)}{d\mu}
\end{array}
\right.
\end{equation}
In terms of a new argument $z=\cos^2(\mu/2)$
(or its continuation on de Sitter space) 
the transversality and eigenvalue conditions imply 
for $\alpha(z)$
\begin{equation}\label{hyper}
z(1-z)\frac{d^2\alpha(z)}{d^2z} + \left[ 4 -8z\right]
\frac{d\alpha(z)}{dz}=(\lambda_{p'} +8)\alpha(z)
\end{equation}
and then for the coefficient functions
\ba\label{app1}
\left\{
\begin{array}{lll}
w_1 & = &
-\frac{6}{5}\left[(\lambda_{p'} +28)z(1-z)- 45/6\right]\alpha (z)
+\frac{6}{20}\left[(\lambda_{p'} +8)z(1-z)(1-2z)\right]
\beta (z)\\
w_2 & = &
\frac{9}{5}\left[(\lambda_{p'} +28)z(1-z)+\frac{20}{3}(1-z)
-\frac{20}{6} \right]\alpha (z)
-\frac{6}{20}\left[(\lambda_{p'} +8)z(1-z)(4-3z)\right]
\beta (z)\\
w_3 & = &
\frac{9}{5}\left[(\lambda_{p'} +28)z(1-z) - 40/6\right]\alpha (z)
-\frac{9}{20}\left[(\lambda_{p'} +8)z(1-z)(1-2z)\right]
\beta (z)
\end{array}
\right.
\ea
with $\lambda_{p'} =(p'^2 +17/4)$.

Notice that equation \ref{hyper} is precisely the hypergeometric 
differential equation, which has a pair of 
independent solutions $\alpha(z)$ and $\alpha(1-z)$ where
\be
\alpha(z)= Q_{p'} \  _2F_{1}(7/2+ip',7/2-ip',4,z)
\ee
$Q_{p'}$ is a constant. 
The solution for $\beta (z)$ follows from equation \ref{bet} and is given by
\begin{equation}
\beta(z) =Q_{p'}\  _2F_{1}(9/2-ip',9/2+ip',5,z).
\end{equation}
We will determine below which solution corresponds to the bitensor
defined by \ref{bit}.

Our discussion so far applies to either $S^4$ or de Sitter space. We
now specialize to the case of $S^4$ and will later obtain results for
de Sitter space by analytic continuation. 
The hypergeometric functions on $S^4$ may be expressed in terms of
Legendre polynomials in $\cos \mu $ (eq. [15.4.19] in \cite{Abram}),
\ba \label{relat}
\left\{
\begin{array}{ll}
\alpha( \mu) & =Q_{p'} \Gamma (4)2^3 (\sin \mu
)^{-3} P^{-3}_{-1/2 +ip'}(- \cos \mu),
\\
\beta( \mu) &=Q_{p'} \Gamma (5) 2^4 (\sin \mu
)^{-4} P^{-4}_{-1/2 +ip'}(- \cos \mu).
\end{array}
\right.
\ea
The solutions for $\alpha(z)$ and $\beta(z)$ are singular at $z=1$
(i.e. for coincident points on $S^4$) for generic values of
$p'$. However, for the values of $p'$ corresponding to the eigenvalues
of the Laplacian on $S^4$, they are regular everywhere on
$S^4$. Similarly, $\alpha(1-z)$ and $\beta(1-z)$ are generically
singular for antipodal points on $S^4$ and regular for these special
values of $p'$. For these special values, $\alpha(z)$ and
$\alpha(1-z)$ are no longer linearly independent but related by
a factor of $(-1)^{(n+1)/2}$ where $n = -2ip' = 7,9,11,\ldots$. 
This follows from the relation (eq.[8.2.3] in \cite{Abram})
\begin{equation} \label{arg}
P^{\mu}_{\nu} (-z) = e^{i\nu \pi } P^{\mu}_{\nu} (z) - \frac{2}{\pi}
e^{-i\mu \pi }\sin [\pi (\nu + \mu )] Q_{\nu}^{\mu} (z),
\end{equation}
where the second term vanishes for $p'=7i/2,9i/2,...$. In fact, the 
hypergeometric series
terminates for these values of $p'$ and the hypergeometric functions
reduce to Gegenbauer polynomials $C^{(7/2)}_{n-7/2}(1-2z )$. We have a
choice between using $\alpha(z)$ and $\alpha(1-z)$ in the bitensor for
these values of $p'$. 
However, to obtain the Lorentzian correlator, we had to
express the discrete sum \ref{cor1} as a contour
integral. Since the Euclidean correlator obeys a differential equation 
with a delta function source at $\mu =0$, we must 
maintain regularity of the integrand at $\mu=\pi$ when extending the bitensor
in the complex $p'$-plane. In other words, for generic $p'$, 
we need to work with the solution $\alpha(z)$, rather then $\alpha
(1-z)$.
We shall therefore choose $\alpha(z)$, since this is the solution that
we will analytically continue.

The above conditions leave the overall normalisation 
of the bitensor undetermined.
To fix the normalisation constant $Q_{p'}$, consider the biscalar
quantity
\ba
g^{ii'}g^{jj'}W_{\ iji'j'}^{(p')}(\mu ) =12 w_{1}^{(p')} -6
w_{2}^{(p')} +24 w_{3}^{(p')}
\ea
In the coincident limit $\Omega \rightarrow \Omega'$ and 
$z \rightarrow 1$ this yields
\begin{equation}
W^{(p')\ ij}_{ij}(\Omega,\Omega)= \sum_{{\cal P}klm}q_{ij}^{(p'){\cal P}klm}
(\Omega) q^{(p'){\cal P}lm\ ij}(\Omega)^{*}
 =-72 \alpha (1).
\end{equation}
Since $F (0)=1$ we have $\alpha (1)= Q_{p'}(-1)^{(1+n)/2}$.
By integrating over the four-sphere and using the normalization condition
\ref{norm} on the tensor harmonics one obtains, for $n = -2ip' = 7,9,11,\ldots$
\be
 Q_{p'} = \frac{ip'(4p'^2 +25)}{48 \pi^2 (-1)^{(1+n)/2}}=
\frac{p'(4p'^2 +25)}{48 \pi^2 \sinh p' \pi}. 
\ee

We conclude that
the properties of the bitensor appearing in the tensor correlator
completely determine its form.
Notice that in terms of the label $p'$ we have obtained a
unified functional description of the bitensor
on $S^4$ and de Sitter space. However, its explicit form
is very different in the two cases because the label $p'$ takes on
different values. It is precisely this description
that has enabled us in section \ref{sec:CFT} to analytically
continue the correlator from the Euclidean instanton
into de Sitter space without Fourier decomposing it.
We shall conclude this Appendix by describing in detail the
subtleties of this analytic continuation at the level of the bitensor.

To perform the continuation to de Sitter space 
we note that the geodesic separation $\mu$ on $S^4$
continues to $\pi /2 -it$, so $z=\frac{1}{2}(1+i \sinh t )$ on de Sitter
space.
The continuation of the hypergeometric functions (\ref{relat}) yields
\ba \label{app3} 
\left\{
\begin{array}{ll}
\alpha(z) & =\Gamma (4)2^3 (\cosh t
)^{-3} P^{-3}_{-1/2 +ip'}(-i \sinh t),
\\
\beta(z) &=\Gamma (5) 2^4 (\cosh t
)^{-4} P^{-4}_{-1/2 +ip'}(-i \sinh t).
\end{array}
\right.
\ea

However, an extra subtlety arises if one wants to
identify the continued bitensor with the usual sum of
tensor harmonics on de Sitter space.
In particular, in order for the bitensor to 
correspond to the usual sum of rank-two tensor harmonics on the real 
$p'$-axis, one must choose the second solution $\alpha (1-z)$ 
to the hypergeometric equation, rather then $\alpha (z)$ that enters
in the continued bitensor.
This is easily seen by performing the continuation on the completeness
relation (equation \ref{com}), which should continue to an integral over $p'$
from $0$ to $\infty$ of the Lorentzian bitensor, defined as the sum
(\ref{bit}) over the degenerate tensor harmonics on de Sitter space.
Writing (\ref{com}) as a contour integral and continuing to Lorentzian
de Sitter space yields
\begin{equation}\label{com2}
g^{-{1\over 2}}
\delta^{ij}_{\ \ i'j'}(x - x') = \int_{-\infty}^{+\infty} 
dp' \tanh p' \pi W^{\ ij}_{(p')\ i'j'}(\mu(x,x')).
\end{equation}
Clearly this is not the correct completeness relation according to the
equivalent definition (\ref{bit}) of the bitensor on de Sitter
space. But let us relate the continued bitensor in (\ref{com2}) to
the independent bitensor in which the solutions
$\alpha (1-z)$ enter.
This can be done by
applying (\ref{arg}) to the Legendre polynomials 
in (\ref{app3}). By closing the contour in the upper half $p'$-plane, 
one sees there is no contribution to the integral (and indeed to the 
tensor correlator!)
from the second term in equation \ref{arg}, because its prefactor 
cancels the $\cosh^{-1} (p'\pi )$-factor in (\ref{com2}),
making the integrand analytic in the upper half $p'$-plane (up to
gauge modes).
Hence, under the integral both solutions are simply related
by the factor $ie^{p\pi}$. In addition one needs to
extract the $\sinh^{-1} p' \pi$-factor\footnote{Remember that
$Q_{p'}$ gained the factor $\sinh^{-1} p' \pi $ because we have chosen
the solution $\alpha (z)$ on the four sphere. The correct
normalisation constant for the independent bitensor, obtained from
the normalisation condition on the tensor harmonics, is then 
$\tilde Q_{p'} = \sinh p' \pi Q_{p'}$.}
from $Q_{p'}$.
The completeness relation then becomes,
\begin{equation}\label{com3}
g^{-{1\over 2}}
\delta^{ij}_{\ \ i'j'}(x - x') = \int_{0}^{+\infty} 
dp' W^{\ ij}_{L(p')\ i'j'}(\mu(x,x')),
\end{equation}
and the hypergeometric functions $\alpha (1-z)$ and $\beta (1-z)$
that constitute
the final bitensor $ W^{\ ij}_{L(p')\ i'j'}(\mu(x,x'))$
are given by
\ba \label{app3a}
\left\{
\begin{array}{ll}
\alpha(1-z) & =\tilde Q_{p'}\Gamma (4)2^3 (\cosh t
)^{-3} P^{-3}_{-1/2 +ip'}(i \sinh t),
\\
\beta(1-z) &=\tilde Q_{p'}\Gamma (5) 2^4 (\cosh t
)^{-4} P^{-4}_{-1/2 +ip'}(i \sinh t),
\end{array}
\right.
\ea
with $\tilde Q_{p'} = p'(4p'^2 +25)/48 \pi^2 $.

On the real $p'$-axis, $W_{\ iji'j'}^{L(p')}(\mu)$ equals the sum 
(\ref{bit}) of the degenerate rank-two tensor harmonics
on closed de Sitter space with eigenvalue $\lambda_{p'}=(p'^{2} + 17/4
)$ of the Laplacian.

\end{document}